\documentclass[sigconf,authorversion]{acmart} 

\acmConference[EASE 2025]{The 29th International Conference on Evaluation and Assessment in Software Engineering}{17–20 June, 2025}{Istanbul, Türkiye}

\usepackage{fancyhdr}
\AtBeginDocument{%
    \addtolength{\footskip}{2.0\baselineskip}%
    \fancyfoot[C]{\textit{Preprint --- do not distribute.}}%
}

\AtBeginDocument{%
  }

\acmDOI{}

\begin{document}
\title[SecCityVR]{SecCityVR: Visualization and Collaborative Exploration of Software Vulnerabilities in Virtual Reality}

\author{Dennis Wüppelmann}
\email{dennis.wueppelmann@uni-paderborn.de}
\orcid{0009-0008-8979-1457}
\affiliation{%
  \institution{Paderborn University}
  \city{Paderborn}
  \country{Germany}
}

\author{Enes Yigitbas}
\email{enes.yigitbas@uni-paderborn.de}
\orcid{0000-0002-5967-833X}
\affiliation{%
  \institution{Paderborn University}
  \city{Paderborn}
  \country{Germany}
}

\begin{abstract}
Security vulnerabilities in software systems represent significant risks as potential entry points for malicious attacks. Traditional dashboards that display the results of static analysis security testing often use 2D or 3D visualizations, which tend to lack the spatial details required to effectively reveal issues such as the propagation of vulnerabilities across the codebase or the appearance of concurrent vulnerabilities. Additionally, most reporting solutions only treat the analysis results as an artifact that can be reviewed or edited asynchronously by developers, limiting real-time, collaborative exploration. To the best of our knowledge, no VR-based approach exists for the visualization and interactive exploration of software security vulnerabilities. Addressing these challenges, the virtual reality (VR) environment \emph{SecCityVR} was developed as a proof-of-concept implementation that employs the code city metaphor within VR to visualize software security vulnerabilities as colored building floors inside the surrounding virtual city. By integrating the application's call graph, vulnerabilities are contextualized within related software components. \emph{SecCityVR} supports multi-user collaboration and interactive exploration. It provides explanations and mitigations for detected issues. A user study comparing \emph{SecCityVR} with the traditional dashboard \emph{find-sec-bugs} showed the VR approach provided a favorable experience, with higher usability (SUS score of 70 to 58.75), lower temporal demand (TLX subscale of 10 to 42.5), and significantly lower frustration (TLX subscale of 17.5 to 52.5) despite having longer task completion times. This paper and its results contribute to the fields of collaborative and secure software engineering, as well as software visualization. It provides a new application of VR code cities to visualize security vulnerabilities, as well as a novel environment for security audits using collaborative and immersive technologies.
\end{abstract}




\maketitle
\section{Introduction}
Security vulnerabilities are a major risk in software systems, often exploited as entry points for attacks. According to the U.S. Department of Homeland Security, 90\% of security incidents are due to software defects~\cite{Morgan.02.09.2015}. There are multiple ways to mitigate the risks of security vulnerabilities, like spreading awareness, embedding security checks into the development process, or conducting active security audits. A common place for awareness is the OWASP Top Ten document, which represents a list of the most critical security risks to web applications~\cite{OWASP10.12.06.2023}. Embedding security into the development process often involves tools like Static Analysis Security Testing (SAST), which detects issues without executing the code. SAST can be applied early—even before compilation—and does not require a complete implementation~\cite{MSDL}. A variety of tools support this, including SpotBugs\footnote{https://spotbugs.github.io/}, SonarQube\footnote{https://www.sonarsource.com/}, and CogniCrypt\footnote{https://eclipse.dev/cognicrypt/}, each targeting different aspects like code quality, style, and security vulnerabilities.

But with all the development in making the tools increasingly more precise and faster, their results are displayed in overloaded dashboards or aggregations without a connection to the related source code. In these bloated reports, it is hard to grasp the kind of vulnerabilities that are present~\cite{9629428}. When visualizing code to understand the security impact of a concrete vulnerability, these aggregations are ineffective due to their neglect of most of the codebase's hierarchy. Two problem statements are derived from this: 
\begin{enumerate}
    \item[$P_1$] Aggregated visualizations of SAST results do not show detected vulnerabilities in the context of the whole code base.
    \item[$P_2$] Aggregated visualizations of SAST results do not show the possible connections of vulnerable software components to other components.
\end{enumerate}A different way to visualize source code has been studied in recent years. By the use of metaphors like solar systems~\cite{graham2004solar}, islands~\cite{Island_Metaphor}, or cities~\cite{Wettel.2007}, it is possible to escape the 2D visualizations and visualize the code in a more meaningful way. A commonly used 3D representation is code cities. Using a city metaphor to represent a code base helps the user to comprehend a code base by giving them a sense of locality~\cite{9604828}. It provides the auditor with a tool to incrementally explore the code, just like they would visit a new city. To create the city, each building represents a single software artifact, e.g., a class. The appearance (height, size, color, texture) of a building can be mapped to metrics suiting the use case of the visualization~\cite{Wettel.2007}. There have been several studies to explore the application of VR in the field of software visualization and collaborative work on artifacts. Experiments comparing the code city metaphor representation in VR with an on-screen representation indicate that VR is a viable means for software visualization~\cite{ROMANO201992}.

Developers express a need for a way to easily communicate and collaborate when using SAST tools if they work as a team. Currently, if there are collaboration features, they provide asynchronous communication through comments or similar review functionalities~\cite{6606613}. Furthermore, the engagement of developers could be improved through the gamification of software engineering tasks with an interactive software visualization tool in VR~\cite{8094470}. This results in the third problem statement:
\begin{enumerate}
    \item[$P_3$] Aggregated visualizations of SAST results are not collaborative and engaging.
\end{enumerate}
To address the problems, a VR environment named \emph{SecCityVR} was created. This paper and the development of \emph{SecCityVR} are guided by the example scenario of a virtual security audit after a vulnerability analysis, where a security professional reviews a Java software repository. The audit focuses on finding the OWASP Top Ten and other known security issues. Together with a developer, they can explore and get an overview of all issues or focus on a high-priority vulnerability and discuss solutions. Additionally, the less specialized developers can use the provided information to learn about different categories of vulnerabilities, the fundamental problems behind them, and ways to prevent them. This leads to the research questions:
\begin{enumerate}
  \item[$RQ_1$] How can software vulnerabilities and their impacts on different system components be effectively visualized using VR and the code city metaphor?
  \item[$RQ_2$] To what extent does visualizing software vulnerabilities in a VR code city aid developers in identifying and exploring the affected system components?
\end{enumerate}This paper examines these research questions by conceptually designing, implementing, and evaluating \emph{SecCityVR} as a proof-of-concept VR environment. It provides a new application of VR code cities to visualize security vulnerabilities and a novel environment for security audits using immersive technologies. Through the usage of collaborative features, together with study participants, we examined a possible multi-user application of this environment. The user study results highlight the benefits of a VR approach, including improved usability, reduced cognitive load, and lower frustration levels, which outweigh the longer task completion times and have significant implications for collaborative and secure software engineering, as well as software visualization. We recommend exploring the novel application of VR in industry software projects to investigate security issues, as well as in academic research to facilitate the communication of findings, algorithms, and techniques.

The sections are structured as follows: Section 2 reviews related work, exploring static analysis, code cities, and their VR applications. Section 3 details the design and implementation of \emph{SecCityVR}. Section 4 outlines the user study, including its design, execution, and results. Finally, Section 5 presents conclusions, the added scientific value, and potential future research.

\section{Related Work}
\label{ch:related_work}
This section gives a structured review of the work done in related fields and compares them based on the research questions. They are rated concerning the use of VR, collaboration, and visualization of vulnerability impact. The Section~\ref{sec:related_work:static_analysis} highlights typical visualization approaches from static software analysis. Section~\ref{sec:related_work:code_cities} provides an overview of different code city approaches, including 3D and VR solutions. Section~\ref{sec:related_work:discussion} summarizes and discusses the reviewed solutions.

\subsection{Static Analysis}
\label{sec:related_work:static_analysis}
An example reference for a common open-source tool that easily integrates into most workflows and detects the OWASP Top Ten is SonarQube\footnote{https://www.sonarsource.com/products/sonarqube/}. Apart from the aggregated dashboard view, it is possible to view every vulnerability or code smell on the detected line of the source code, which supports the demand from the interviewed developers in~\cite{6606613}, that: 'an analysis output should be a 'slice' that shows what the problem is and what else could be affected to more quickly assess what is or is not important'. While SonarQube supports over 30 programming languages, there are tools and frameworks optimized for specific languages like FlowDroid~\cite{tuprints5937}, which provides static data flow analysis for Android applications. More special use cases are supported by optimized tools like the misuse of cryptographic APIs (with CogniCrypt~\cite{8115707}) or automated detection of over-privileged policies in an AWS infrastructure (with CodeShield\footnote{https://codeshield.io}). Notable examples of foundational frameworks that operate on intermediate representations of programming languages include PhASAR~\cite{PhASAR} and SootUp~\cite{SootUp}, which serve as a basis for developing specialized and advanced tools. The tools available can be integrated into the developer workflow to varying degrees, but there are also solutions designed specifically for this, such as GitHub's use of CodeQl in GitHub Actions\footnote{https://docs.github.com/en/code-security}.
If the analysis is security related, \citeauthor{9629428} summarized that the visualization usually is graph or treemap-based~\cite{9629428}.

\subsection{Code Cities}
\label{sec:related_work:code_cities}
\citeauthor{Wettel.2007} took the idea of a treemap algorithm and used it as a layout algorithm to visualize software code as a city where each class is represented by a building in the city. In their created tool \emph{CodeCity}~\cite{Wettel.2007}, source code metrics are mapped onto the size and type of buildings. They used the number of methods as building height and the number of attributes as building base size. The buildings are organized in rectangular districts which represent a package containing multiple classes. The chosen level of detail (building = class) is due to the fact that classes are the easiest orientation point for developers. 

\citeauthor{7739576} support their idea that a treemap maintains the codebase's hierarchy to which developers are accustomed to maintaining closeness to the programming environment~\cite{7739576}. Although their 3D visualization is displayed on a computer screen without collaboration features or the integration of vulnerabilities, CodeCity is groundlaying and there have been many improvements and experiments based on the work and ideas of \citeauthor{Wettel.2007}.

\subsubsection*{1) Explored code city solutions}
The research of \citeauthor{dep_viz_city_layout}, as an example, adapted the city metaphor to visualize dependencies in software by changing the metrics to building height representing the number of incoming dependencies and the square area representing the number of outgoing dependencies~\cite{dep_viz_city_layout}. For a better representation, they adapted an intuitive layout that is based on a layered graph drawing as encoding for dependencies, which reduces arcs, and drawn ones usually indicate architecture violations. The arcs in their tool can also be used to trace a root cause analysis. Focusing on software architecture, no statically detected software vulnerabilities are highlighted but the authors hint that collaboration should be integrated into their solution.

Instead of simply coloring the rendered building, \citeauthor{evo_spaces_architecture} increased the level of detail of the city in their tool EvoSpaces~\cite{evo_spaces_architecture}, by classifying buildings into four categories: city halls, houses, apartment blocks, and skyscrapers. Each class received a different model and texture, to better represent the class in the city environment. The non-collaborative tool includes a trace analysis feature after code execution in a dynamic analysis fashion. The visualization approach uses a night mode to realistically depict the trace, with illuminated buildings indicating the ones in which 'the people are working' (the executed methods). This facilitates partial exploration of the trace, but not the vulnerability itself.

There have been attempts to combine both SAST tools and the city metaphor. For example, the already mentioned tool SonarQube was extended by a plugin called SoftVis3D\footnote{https://softvis3d.com/} to render a 3D code city based on the SonarQube results. But these only represent a static set of metrics like the mapping of a class' number of vulnerabilities to its color, there is no view of affected code for a vulnerability, its embedding in a call graph, or VR support. 

The described tools and experiments above only visualized the code city in a graphical 3D rendering. There have been attempts to increase the immersive power of the city metaphor implementation even more. \citeauthor{code_metropolis} used the video game Minecraft to visualize the results of a static analysis users could run in the background while developing in their IDE. Afterward, the user could run or fly around in their generated city inside the common game environment~\cite{code_metropolis}. This enabled multiple users to collaboratively inspect the generated city but neither a VR environment nor security vulnerabilities were explored. 

\subsubsection*{2) Virtual Reality Solutions}
A different take on gamification explored \citeauthor{8094470} The authors created an immersive VR environment to boost developer engagement in software comprehension tasks. In a case study, they found that developers felt curious, immersed, in control, excited, and challenged. The perceived time passed faster than in reality, and therefore developers were willing to spend more time using the tool to solve software engineering tasks. Although no static analysis results were displayed in the city, authors envisioned that in the future multiple users should use VR code cities collaboratively~\cite{8094470}.

\citeauthor{koschke2021modeling} developed the SEE-Tool, a multi-functional software visualization based on the software city metaphor. In an experiment they found, that code cities are better suited to get a quick overview of the code smells of a software, whereas tabular representations are better suited to analyze code smells in more detail. Their software SEE supports developers in metric visualization, monitoring the evolution, debugging tasks, and an architecture overview. It provides a digital and collaborative VR environment around different city metaphor visualizations with multiple users. Their city implementation supports different layout algorithms: Circular Balloon, Circle Packing, Rectangle Packing, Treemap, and an EvoStreets layout. They compared their built tool with a professional dashboard application as a baseline, specifically designed for the presentation of code-smell data~\cite{Galperin.2022, koschke2021modeling}.

To provide the possibility to represent static metrics and method coupling, dynamic tracing of executed code \citeauthor{8004366} extended the layout and skyscraper rendering by software metrics \& source code modifications at the granularity of methods. The tool also supports visualization of the software evolution by aggregating the contributions of authors~\cite{8004366}. Through dynamic tracing, a given vulnerability's impact at runtime could be investigated but no further support for security vulnerabilities or collaboration is given.
\subsection{Discussion}
\label{sec:related_work:discussion}
Explored use cases and domains of code cities, in the reviewed literature, include: understanding a codebase, performance visualization, debugging, and detecting architectural flaws. Most cities map classes to buildings, but detailed granularity (e.g., method-level views) and multi-codebase integration are rare. Treemap-based layouts dominate, except for studies focused on improved city design. To the best of our knowledge, no VR-based approach exists for the visualization and interactive exploration of software security vulnerabilities. 

The review of the related work ends with the conclusion that the tool closest to the research questions is the SEE tool created by \citeauthor{koschke2021modeling}. The main difference is the overall setting of the VR environment. In SEE multiple code cities, each with a different purpose of visualization, are placed on different tables in one virtual room where multiple users can walk through the room and investigate the cities as artifacts in the room. In \emph{SecCityVR}, our code city optimized for vulnerability visualization, one city is chosen as the surrounding environment which can be explored by flying or walking on the ground. It is focused on one use case: the (collaborative) exploration of visualized software vulnerabilities.

\section{SecCityVR Environment}
\label{ch:implementation}
\begin{figure}[hb]
  \centering
  \includegraphics[width=\columnwidth]{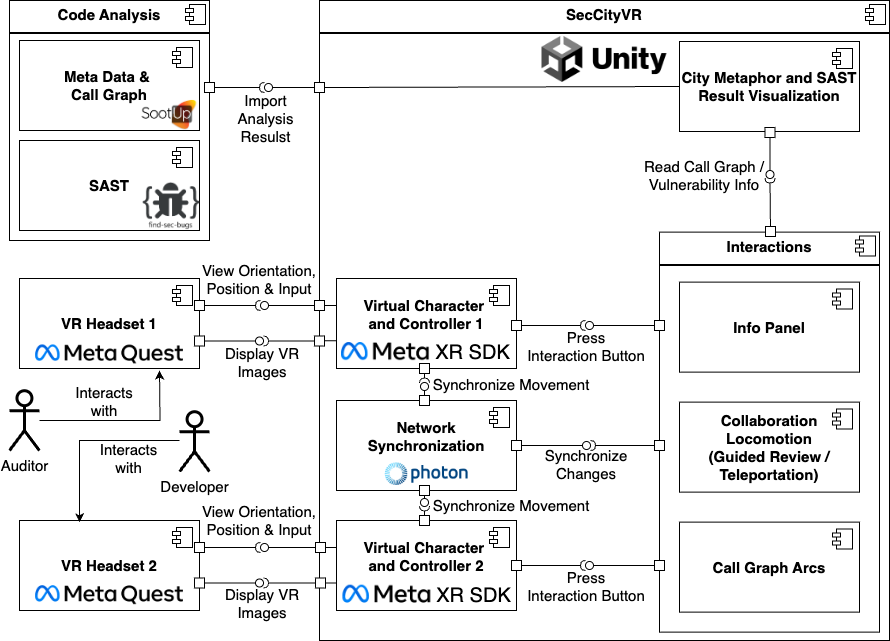}
  \caption{Architecture Overview of \emph{SecCityVR}}
  \label{fig:ComponentDiagramSecCityVR}
  \Description{UML Component Diagram of the \emph{SecCityVR} System}
\end{figure}

In Figure~\ref{fig:ComponentDiagramSecCityVR}, an architectural overview of the developed system is given. The whole system is split into two parts. First, the groundlaying \emph{Code Analysis}, consisting of the \emph{SAST}, \emph{Meta Data \& Call Graph} analyses, which is further described in Section~\ref{sec:impl:analysis}. Second, the \emph{SecCityVR} environment, representing the main component, uses the analysis results as input to render the city and display the vulnerabilities (described in Section~\ref{sec:impl:viz}). It is an environment based on the \emph{Unity} engine that can be explored using a \emph{Meta Quest VR Headset}. Through \emph{Interactions} with their controllers, users can move through the city, select methods, read about their security issues, and follow the \emph{Call Graph Arcs} of vulnerable methods (described in Section~\ref{sec:impl:interactions}). In addition to that, features for \emph{Collaboration} and \emph{Network Synchronization} are included to investigate or explore the code city with multiple users (described in Section~\ref{sec:impl:collab}).
The \emph{SecCityVR} application is available online\footnote{https://github.com/dewue/SecCityVR} as an open-source \emph{Unity} project, along with a demonstration video\footnote{https://uni-paderborn.sciebo.de/s/jvNvVwW8QzYmOBv} showcasing the VR application and its integrated tutorial.
\subsection{Code Analysis}
\label{sec:impl:analysis}
Two independent static analyses are part of the \emph{Code Analysis} component. First, a SAST tool that detects the OWASP Top Ten security vulnerabilities of a Java Archive (JAR). Second, an analysis extracting the package hierarchy and metadata like lines of code as well as the call graph of the JAR.

We utilize Find Security Bugs (find-sec-bugs)\footnote{https://find-sec-bugs.github.io/} as our static analysis security testing tool, which extends SpotBugs with security audits and offers a command-line interface (CLI) for execution. The executed command generates a .xml file containing the extracted vulnerabilities from a specified JAR file. It contains all security issues listed per method as well as a description of the issue and often a solution and further resources about it. 

The second textual artifact, capturing the package and call hierarchies, is generated by a custom tool developed for this work. In the execution of this, the SootUp framework is first used to construct the call graph of the given JAR file. The call graph construction is executed using SootUp's implementation of the Rapid Type Analysis algorithm to receive a call graph with possibly instantiated edges. In the next step, with the created graph as input for querying, the JAR is analyzed regarding its internal packages and their hierarchy.  The result is exported as a .json file, serving as input for rendering the code city, which is later augmented with detected vulnerabilities.


\subsection{City Metaphor and SAST Visualization}
\label{sec:impl:viz}
Processing of the analysis results starts with importing the generated textual artifacts into the \emph{Unity} project. They are stored as assets in the project and loaded as inputs to the main visualization at the start of \emph{SecCityVR}. The chosen mapping of concepts of the code base to elements of the visualization is to represent classes as buildings grouped by their package, displayed as a platform where the buildings are placed. Methods appear as floors of the buildings but are not rendered by default, to save resources. If a method contains a detected vulnerability, the corresponding part of the building gets colored according to the priority of the issue. To emphasize the methods even further, they are rendered slightly wider than their class building. Every method's size and position are rendered according to its line of code metric and its start/end position inside its class. Such a visualization approach will aid the goal of the environment, which is supporting the exploration and not the specific investigation (for which a more detailed approach, like using methods as buildings, could be better).

The next step in the direction of a rendered city is the placement of these buildings. Different layout algorithms were explored in the past, from simple chessboard placing to more code structure representing layouts like treemap-based, hierarchical ones, or sophisticated altered layouts optimized for one use case like~\cite{dep_viz_city_layout}. A quadratic baseplate is generated using the repository’s total lines of code. The city layout is created with a squarified treemap algorithm that subdivides this baseplate into rectangles representing packages, sized by their line count. Our implementation builds on the work of \citeauthor{TreemappingC}\cite{TreemappingC} and \citeauthor{bruls2000squarified}\cite{bruls2000squarified}, with adaptations for floating-point support, improved variable typing, and a custom output format for code city rendering. The algorithm processes the package hierarchy level by level, aiming to keep buildings as square as possible. The package tree is processed level by level, and an outline of the recursive process looks like this:

\begin{enumerate}
    \item First algorithm run 
    \begin{itemize}
        \item Package rectangles as 'districts'
        \item Adds spacing for 'streets' 
    \end{itemize}
    \item Second algorithm run 
    \begin{itemize}
        \item Classes as 'buildings' inside their package districts
        \item Extrude buildings to the matching lines of code
    \end{itemize}
\end{enumerate}The third step is the emphasizing of methods in the city metaphor according to their severity. Two visual effects were implemented to support the rendering of them as a floor of their class building. First, methods receive a color based on the highest vulnerability found in their code. This aligns with the original color coding from the analysis tool SpotBugs, where bugs are categorized and colored by priority: high (red), medium (orange), low (green), or info/experimental (blue). Second, the width of the method is wider than the cube of its building. The position and size of a method floor are calculated with the start and end line of code of the method in their class.

Based on the generated buildings, the only missing information for an accurate representation of the repository is the information about connected methods. For this, the constructed call graph from the JAR analysis is used to show the connection of tainted parts of the software. The call graph edges between the methods, pictured in Figure~\ref{fig:SecCityExample}, are represented as arcs above the buildings, connecting the rendered methods. To reduce the number of arcs drawn at one moment of time, the city starts with none of the methods connected visually. If the user triggers the visualization of incoming and outgoing call graph edges, for a method with vulnerabilities, connected methods get enabled. They are visualized with the same two visual highlights vulnerable methods receive, blue color and a slightly wider width than the parent building. The arc gets colored from blue to white through the application of a color gradient. In the case that both methods are contained in the same class, the arc is not drawn above the class building, but on the side of a building.

In modern software applications, most functionality is built on frameworks. These dependencies are usually not all maintained by the author of the original code. Additionally, they often include security issues themselves. To easily uncover vulnerabilities in code the main application depends on, the virtual code city must include buildings and districts representing the dependencies' code as well as the main code of the repository. Packages containing the main application code should be visually distinguishable from their dependencies so that developers can identify their code and focus on issues in the code they own. For this, a coloring of the rendered package platform in a color not utilized for transporting any issue information is used, which is magenta.

\subsection{Interactions}
\label{sec:impl:interactions}
\begin{figure}[b]
  \centering
  \includegraphics[trim={2cm 0cm 10cm 26cm},clip,width=0.8\columnwidth]{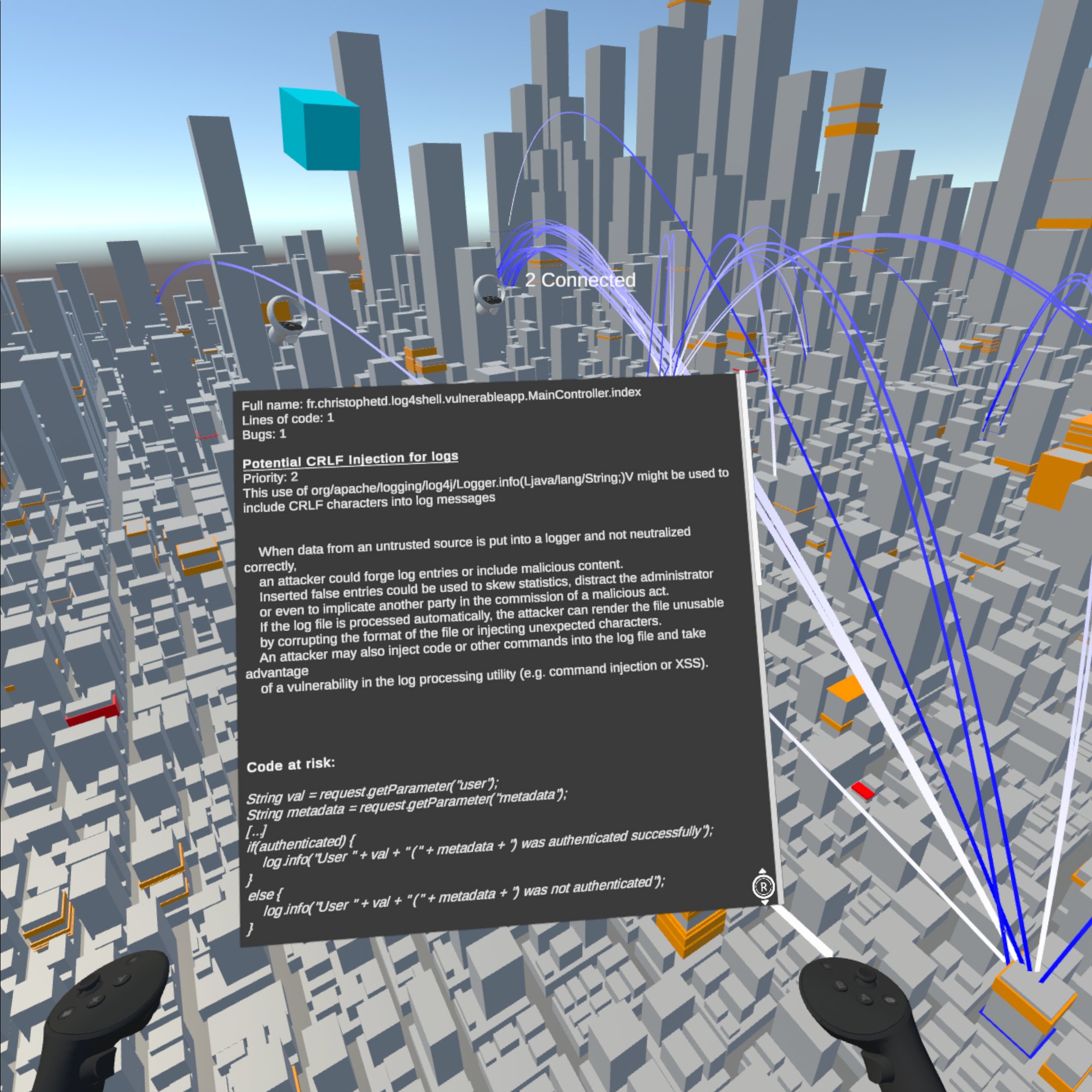}
  \caption{In-app screenshot of the \emph{Info Panel}}
  \label{fig:SecCityExamplePanel}
  \Description{VR Environment with the city in the background and the info panel showing a vulnerability description}
\end{figure}
The described system so far fully represents the code base using the code city metaphor, preserving the package-to-class hierarchy and marking methods with security issues as zones of interest. Interaction within the virtual world is enabled via official \emph{Meta Quest} controllers. Users can press the menu button to display a control guide and color legend. A digital replica of the controllers is also rendered in VR, synchronized with the user's hand movements. 
\begin{figure}[ht]
  \centering
  \includegraphics[trim={5cm 6cm 1cm 5cm},clip,width=0.8\columnwidth]{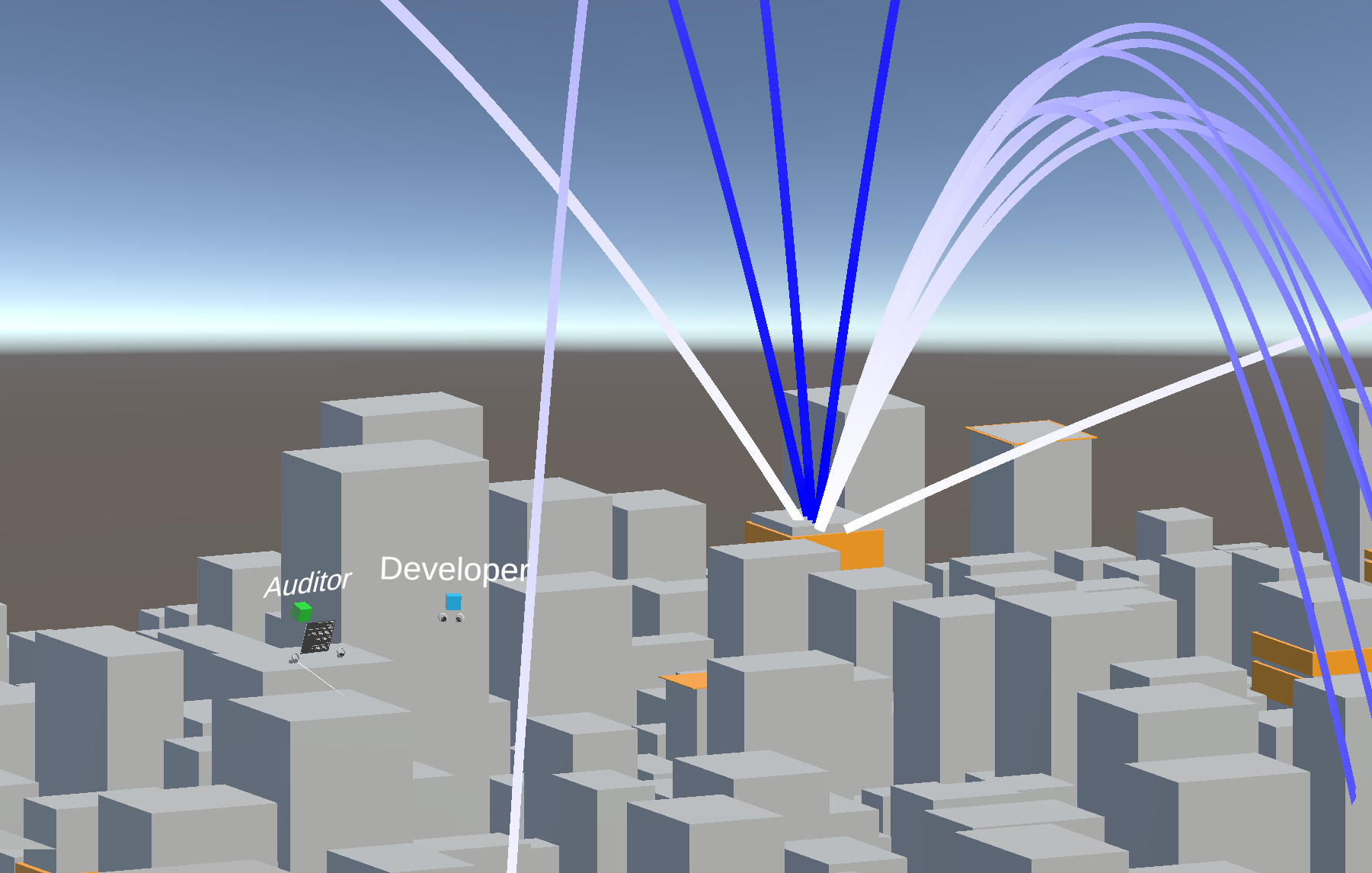}
  \caption{In-app screenshot (with two users as colored cubes)}
  \label{fig:SecCityExample}
  \Description{VR Environment with Auditor and Developer discussing an issue above the code city, the Auditor is reading on the info panel.}
\end{figure}
The first interaction of interest is the locomotion through the world to reach points of interest in the city. In general, the user can move freely through the city. Two kinds of movement are provided for this. Firstly, 'walking' on the streets of the city fully immerses the user in the city and its skyscrapers. This is done by teleportation on the ground level. A user points the controller to a spot on the street and gets teleported to the marked point. Secondly, a flying mode where the user can fly through, above, or around the city. The flying mode confronts the user with less realism, because it is not a known locomotion but delivers a greater possibility for an overview of the city and its parts. Once a user has moved to a point of interest, most of the time, this will be a method, the only visible information is the color-coded priority and the size of the method. To show more details about these points of interest, the selection of the city element is necessary. For selection, the user has to point a controller at a city element and press the mapped trigger button (similar to an abstracted gesture of shooting a gun). A panel will appear to display the name, lines of code, and details about the vulnerabilities in the method. Further, this is called \emph{Info Panel}. To see how a vulnerability is displayed on the panel, view Figure \ref{fig:SecCityExamplePanel}, and to picture how the panel is projected in the environment, see the grey canvas in front of the auditor in Figure~\ref{fig:SecCityExample}. For methods containing multiple vulnerabilities, each with a detailed description, the \emph{Info Panel} includes a scrolling functionality mapped to the joystick of the controller. The last possible information the code city can provide for a method is its embedding in the constructed call graph. After selecting a method, the user can press the \emph{Render Call Graph button} and all incoming and outgoing edges are rendered as arcs from method to method.
\subsection{Collaboration \& Network Synchronization}
\label{sec:impl:collab}
Until now the described environment only combines the two textual artifacts into one rendered 3D visualization, that can be explored. To further enhance the VR solution, collaboration features were added. This allows multiple VR users to connect to the same server over the network and explore the same city. Each of them is displayed with a synchronized cube as a head and two \emph{Meta Quest} controllers as hands to aid the collaboration between multiple users. In Figure~\ref{fig:SecCityExample} the auditor (green cube) and the developer (blue cube) are shown with their virtual characters.
When a user interacts or moves in the city of their local application instance, their interactions and positions are mirrored in the other user's application instance. This is done through the integration of \emph{Photon Fusion}\footnote{https://www.photonengine.com/fusion}, a networking library for \emph{Unity}. While moving through the city, the individuals might lose sight of each other. Therefore, to quickly regroup, with the press of a button, one user can teleport to another user's position. In addition to that, a guided review is possible. With this, as soon as one user enables it, the current user (developer) position gets synchronized with the position of the other user (auditor). Now the developer automatically follows the auditor when they move, this includes flying as well as ground teleportation. While the guided review feature is active the following developer still has control over their head to look around and it is possible to stop following when pressing the \emph{guided review button} again.

To enable users to work collaboratively in the city, different interactions need different executions regarding their collaborative usage. The help and information panels are only used client-sided for one user. This should prevent disturbance while reading through the provided information and should allow users to read at their own speed. The visualization of the call graph edges is different. In this case, each rendered edge and each highlighted connected method should be synchronized with all users to allow collaborative exploration of the same city visualization. As soon as a user presses the \emph{visualize call graph button} for a method, all connected users in this code city session receive the event to trigger the rendering of call edges of the selected method.

\subsection{Scalability \& Integration}
\label{sec:user_study:scalability}
Currently, \emph{SecCityVR} is a standalone visualization tool that renders a code city based on two input files: SAST results and hierarchical JAR metadata. These are generated via a simple CLI command using external tools. If a team prefers a different SAST tool or needs custom data in the visualization, the file loader can be adapted accordingly. As long as reports follow a similar structure and information references a fully qualified method name, the city can display this information in its panels or via custom visualizations. To further improve its integration into software engineering workflows, we propose two directions. First, automating artifact generation via CI/CD would keep the VR city in sync with the latest codebase and analysis. A typical workflow might involve committing code, triggering automated SAST in the pipeline, and then reviewing results collaboratively in \emph{SecCityVR} if vulnerabilities were detected. Second, IDE integration, such as syncing the user's VR position with their IDE cursor (as explored by \citeauthor{code_metropolis}), could provide a seamless transition from IDE to VR and an easier starting point for exploration\cite{code_metropolis}.

\section{Evaluation}
\label{ch:user_study}
Given the initial research questions, a user study was conducted and aimed to evaluate the usability of the visualization in VR with the following hypotheses:
\begin{enumerate}
  \item[$H_1$] Using the code city metaphor for visualizing security-related static analysis results in VR enables users to solve the same tasks as on a tabular dashboard.
  \item[$H_2$] The usability of security issues visualized in VR is at least as good as a tabular dashboard showing the same data.
\end{enumerate}Additionnally, an exploration of the collaborative aspect of \emph{SecCityVR} is done in a multi-user task.

The evaluation follows a between-subjects design where the participants are assigned to use either the baseline \emph{find-sec-bugs} dashboard or \emph{SecCityVR}, and it is split into two experiments. First, a user study for evaluating the exploration features of the VR environment is conducted with a single participant per experiment run (later called main study). Secondly, a user study for testing out the collaboration features is conducted with two participants working together in one experiment run (later called collaboration study). In each of these scenarios, participants have to use their assigned tool to solve given tasks and answer a questionnaire afterward. Each selected task corresponds to a key feature of the environment, and our questionnaire is designed to facilitate a direct comparison between the VR environment and the baseline. Features to be evaluated are the code city metaphor, city layout, issue visualization, issue education, and call graph visualization. As well as the collaboration features with the second task set. The collaborative study setting is designed with one main task in mind: Participants are asked to select two vulnerabilities that, according to their priority criteria, they would like to address first. The tasks for the main study, listed below, are designed to be used for the experiments in the VR group as well as in the baseline group. 

\begin{itemize}
    \item[T1]{(metaphor): \emph{Find the package with most lines of code. Package Name? How many lines of code?}}
    \item[T2]{(vulnerabilities): \emph{Find the package containing the classes of the application package and its vulnerability. Type of Vulnerability? Priority? Solutions?}}                                                          
    \item[T3]{(callgraph traversing): \emph{Find the method 'A' in the class 'XYZ' in the already rendered call graph edges. How many lines of code? Which issue?}}    \item[T4]{(callgraph coloring): \emph{Using the method 'A' what are incoming and outgoing calls? Class and method name as well as the direction of each call?}}
\end{itemize} The baseline dashboard is created using the same analysis that is used to generate the initial .xml artifact containing the security vulnerabilities for visualization. The participants are not allowed to use the browser's search function to keep the baseline similar in features to the VR code city. Participants are also provided the call graph results as the .json text file because the drawing of the call graph exceeded the rendering capabilities. Due to the same reason, the participants are allowed to use the search function of the editor for this file. In the collaborative setting, the same tools and artifacts are used. Participants are provided one laptop with two files each.

\subsection{Procedure}
\begin{table*}[ht]
  \centering
  \begin{minipage}{.4\linewidth}
    \centering
    \begin{tabular}{|c|c|}
      \hline
      Title                    & Count \\\hline
      Student                  & 6     \\\hline
      Scientific Assistant     & 1     \\\hline
      Research Associate       & 1     \\\hline
      Software Developer       & 2     \\\hline
      Software Engineer        & 2     \\\hline
      Senior Software Engineer & 4     \\\hline
      Software Architect       & 1     \\\hline
    \end{tabular}
    \caption{Job Titles}
    \label{tab:job_titles}
  \end{minipage}
  \begin{minipage}{.29\linewidth}
    \centering
    \begin{tabular}{|c|c|}
      \hline
      Experience & Count \\\hline
      <1 year    & 1     \\\hline
      1–3 years  & 3     \\\hline
      4–6 years  & 8     \\\hline
      7–10 years & 4     \\\hline
      10+ years  & 1     \\\hline
    \end{tabular}
    \caption{\\Job Experience}
    \label{tab:job_experience}
  \end{minipage}
  \begin{minipage}{.29\linewidth}
    \centering
    \captionsetup{justification=centering}
    \begin{tabular}{|c|c|}
      \hline
      Experience   & Count \\\hline
      None         & 3     \\\hline
      Beginner     & 6     \\\hline
      Intermediate & 6     \\\hline
      Advanced     & 1     \\\hline
      Expert       & 1     \\\hline
    \end{tabular}
    \caption{\\Security Experience}
    \label{tab:security_experience}
  \end{minipage}
\end{table*}
In this evaluation, a post-study questionnaire after all tasks is used in favor of post-task questionnaires after each task to avoid disturbing the participants' immersion in the VR world. The study starts with a demographic questionnaire consisting of age, gender, academic background, professional experience, VR experience, and experience with SAST. Followed by general instructions and tool-specific instructions before starting the task sets and the integrated VR tutorial. The post-study questionnaire is designed to include several pre-existing and established questionnaires, selected from standardized questionnaires, inspired by those used in related literature's evaluation processes. Perceived dimensions of interest to measure with a focus on evaluating the given hypotheses are usability and the workload while executing the tasks. These are then used to compare the two groups and their assigned tool.

The full post-study questionnaire for the VR group consists of the System Usability Scale (SUS)~\cite{SUS_Questionnaire},  NASA Task Load Index (TLX)~\cite{HART1988139}, Virtual Reality Sickness Questionnaire (VRSQ)~\cite{KIM201866}, Igroup Presence Questionnaire (IPQ)~\cite{IPQ_online}, and additional feedback. The questionnaire and raw results are available online\footnote{https://uni-paderborn.sciebo.de/s/waPW9A03zv1Uu55}. For the dashboard group, only TLX, SUS, and additional feedback are asked. Questionnaires were mainly selected because of fewer questions while being more suited for VR scenarios, to not overload the participants, and to motivate them to participate in the collaboration study. The study was executed with the help of 17 unique participants who received no compensation and volunteered to participate. Participants were recruited from three fields: students/researchers who work with VR, researchers/developers working in secure software engineering research, and software engineers with professional Java experience from an industry company. For the evaluation of the collaborative characteristics of the VR environment, participants were taken from the same pool as the participants for the main study. At the end of each participation, the person was asked if they wanted to partake in the collaboration study. Around half of them agreed and participated in the second part. Only one substitute participant had to join and was not involved in the study setting before, but came from the same pool of initially invited students, researchers, and software engineers. In order to maximize the feedback on the VR environment while minimizing potential learning effects, participants were assigned the opposite tool they had used in the initial study run. This resulted in 8 participants for the second run who were, in pairs of two, assigned to either the baseline dashboard or the VR environment. 

The user studies were conducted in quiet, empty rooms without any distractions. For each experiment, a maximum time slot of one hour was planned, but none of the participants exceeded it. 25 minutes were allocated for them to do all tasks, an additional 15 minutes for instructions, including the VR tutorial, and around 10 minutes for post-study feedback and questionnaires. The final 10 minutes were used as a buffer between participants to reset the used tools. The task time was measured from the moment an instruction was read out loud the first time until the final solution was provided.

Before stating the results, a comment about testing the significance of the data and differences between the studied groups should be made. First, for all reported questionnaires, the Shapiro-Wilk test was done to check the normality. For almost every scenario, it results in a non-significant p-value, which concludes that the study data is not distributed normally. Therefore, the non-parametric Mann-Whitney-U test is used to check for significant differences between the baseline and VR groups. In the following results, each reported p-value belongs to the performed Mann-Whitney-U test. The chosen significance level will be $\alpha = 0.05$, which is a similar procedure to \citeauthor{Galperin.2022}~\cite{Galperin.2022}.

\subsection{Results}
\label{sec:user_study:results}
The main results of the user study are presented in the following section.

\subsubsection*{1) Demographic Data}
All demographic data is reported for all 17 participants, aged 22 to 34 (median 26). Participants have diverse backgrounds, ranging from \emph{Bachelor Students} with no experience to \emph{Senior Software Engineers} and \emph{Software Architects} with 10+ years of experience (see Table \ref{tab:job_titles} and \ref{tab:job_experience}). Professional experience in years is normally distributed, with about half having \emph{4–6 years} and the rest split between \emph{1–3 years} or \emph{7–10 years}, plus outliers below one year and above ten. The self-assessed security-specific experience is skewed lower, with one \emph{advanced} and one \emph{expert}, while most identified as \emph{beginner} (3), \emph{intermediate} (6), or having no experience (3), see Table \ref{tab:security_experience}.

\subsubsection*{2) Efficiency}
\begin{figure}[ht]
  \centering
  \includegraphics[width=\columnwidth]{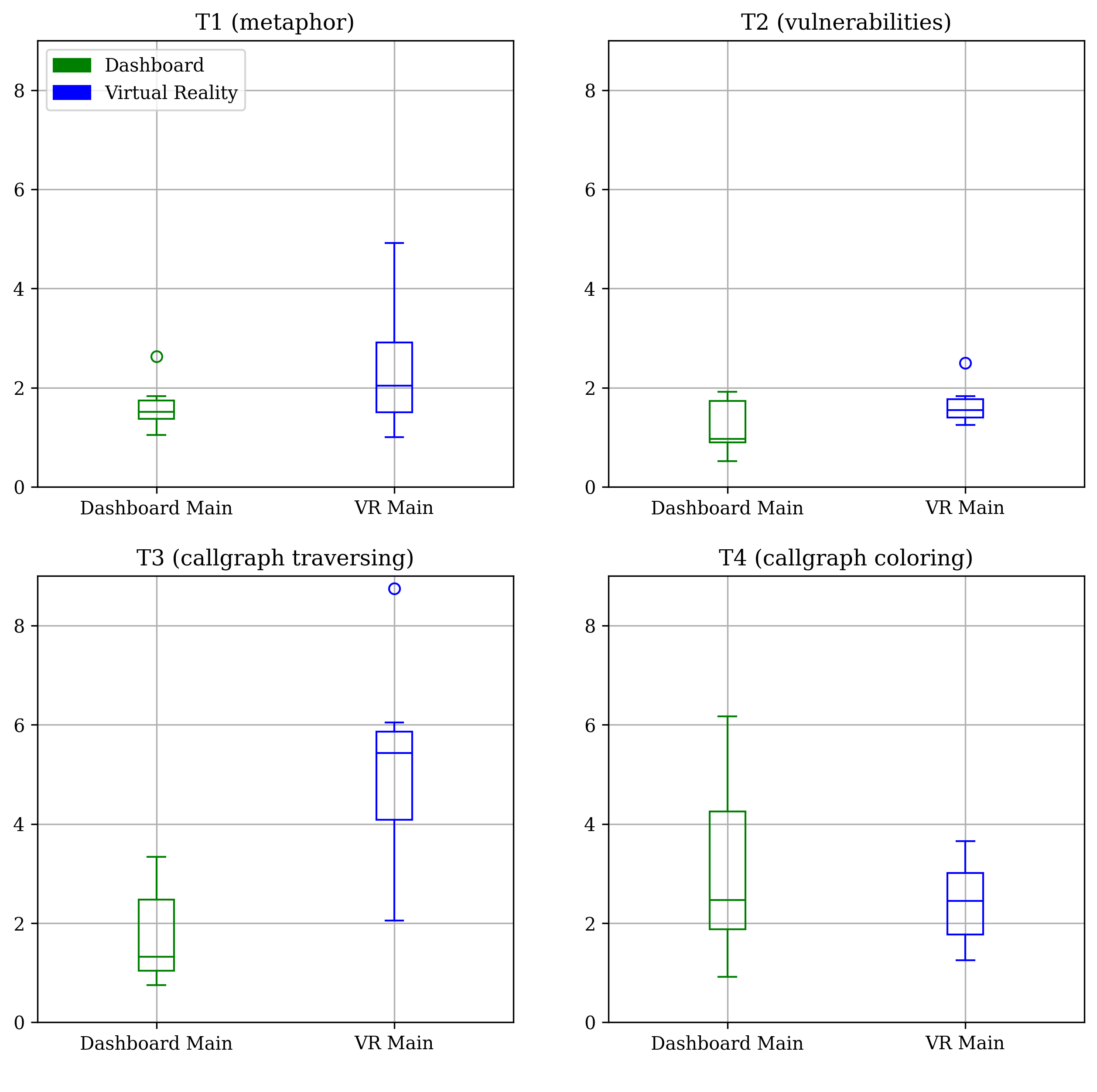}
  \caption{Time per task (in minutes)}
  \label{fig:result_time_tasks_main}
  \Description{Boxplots showing time in minutes per study group, grouped by task}
\end{figure}In Figure~\ref{fig:result_time_tasks_main} the time spent on each task is broken down for each study run of the main study. It can be seen that the participants using the VR environment on average took 04:06 minutes longer for the whole task set than users with the dashboard. Using the Mann-Whitney-U test this overall time spent on tasks reports a significant difference of the dashboard users with $p=0.021^*$. The time measurement only reports the raw task time not accounting for the time in between tasks, therefore not the full time a participant spent in the VR environment. The time per task measurement shows that VR users were slower in all tasks except the fourth one. In the collaborative setting, the main task was limited to 30 minutes. Two teams, one dashboard group, and one VR group both asked to finish this task earlier because they wanted to decide on a final answer. The VR team that aborted took 25:00 minutes, the baseline team selected their issues after 16:10 minutes. While the collaboration tasks were executed, the participants were observed and their strategy was noted. Additionally, their provided feedback about features while thinking out loud was noted and consolidated later.

\subsubsection*{3) Usability}
The first analyzed dimension, with the questionnaires' results, is usability measured by the score calculated from the SUS questionnaire. The median SUS score of all VR users is 70.0 while the median of all dashboard users is rated 58.75. \citeauthor{SUS_Benchmarks} state in their tips for usability practitioners that a score of 68 could be used as a benchmark for average usability, which rates the VR environment with above average usability and the dashboard usability below average usability rating~\cite{SUS_Benchmarks}. However, the Mann-Whitney-U test resulted in no reported significance $p = 0.14$.
\begin{figure}[ht]
  \centering
  \includegraphics[width=0.45\textwidth]{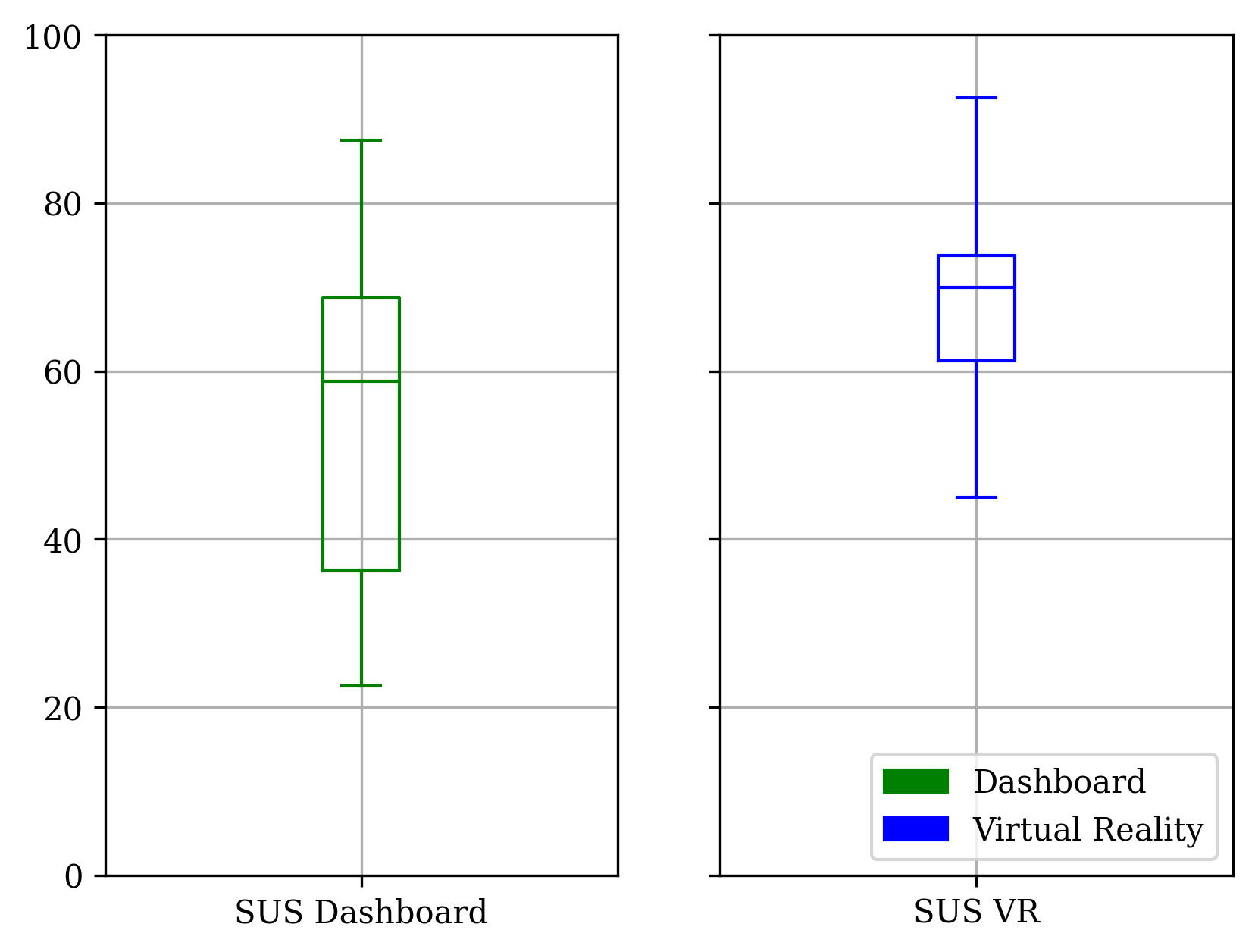}
  \caption{SUS values comparison of both study runs }
  \Description{Two boxplots of SUS Results}
  \label{fig:result_sus_box}
\end{figure}

\subsubsection*{4) Task Load}
The perceived task load of participants is assessed with the NASA TLX questionnaire. Unlike for the SUS, the results of the two studies are not combined due to the different task sets. The reported median TLX of the main study is similar between both tools except for two questions. Firstly, for the temporal demand of the dashboard task with a median value of 42.5 the VR only received 10.0 ($p=0.061$). Secondly, on the frustration dimension dashboard, users rated it with a median of 52.5 while VR users only rated their frustration with a median of 17.5 ($p=0.03^*$). In the collaboration study, these tendencies show again with a median rating on temporal demand of 77.5 to 25 ($p=0.029^*$) and frustration of 65 to 32.5 ($p=0.2$). Additionally, in this study run the dimensions of mental demand (75 Dashboard, 52.5 VR, $p=0.303$), physical demand (12.5 Dashboard, 65 VR, $p=0.029$), and performance (72.5 Dashboard, 52.5 VR, $p=0.2^*$) differ notable, even though out of these only the physical demand reports with significance.

\subsubsection*{5) Presence}
With the IPQ questionnaire presence, as the general sense of being in the virtual environment, is collected. Apart from immersion as an attribute of technology, the measured presence is the subjective experience of participants. Participants answered 14 questions, on a 7-point Likert scale valued from 0 to 6 in the evaluation. The questionnaire measures three subscales that combine into one \emph{Presence Score}, the higher the better. The final \emph{Presence Score} of the IPQ has a median of 3.21. Reviewing the raw results this is rated above the possible average rating of 3 on the questionnaire scale from 0-6. The median results are 3.5 on the \emph{Spatial Presence} subscale, 3.375 on the \emph{Involvement} subscale, and for the \emph{Experienced Realism} subscale 2.25.
\begin{figure}[hb]
  \centering
  \includegraphics[width=0.45\textwidth]{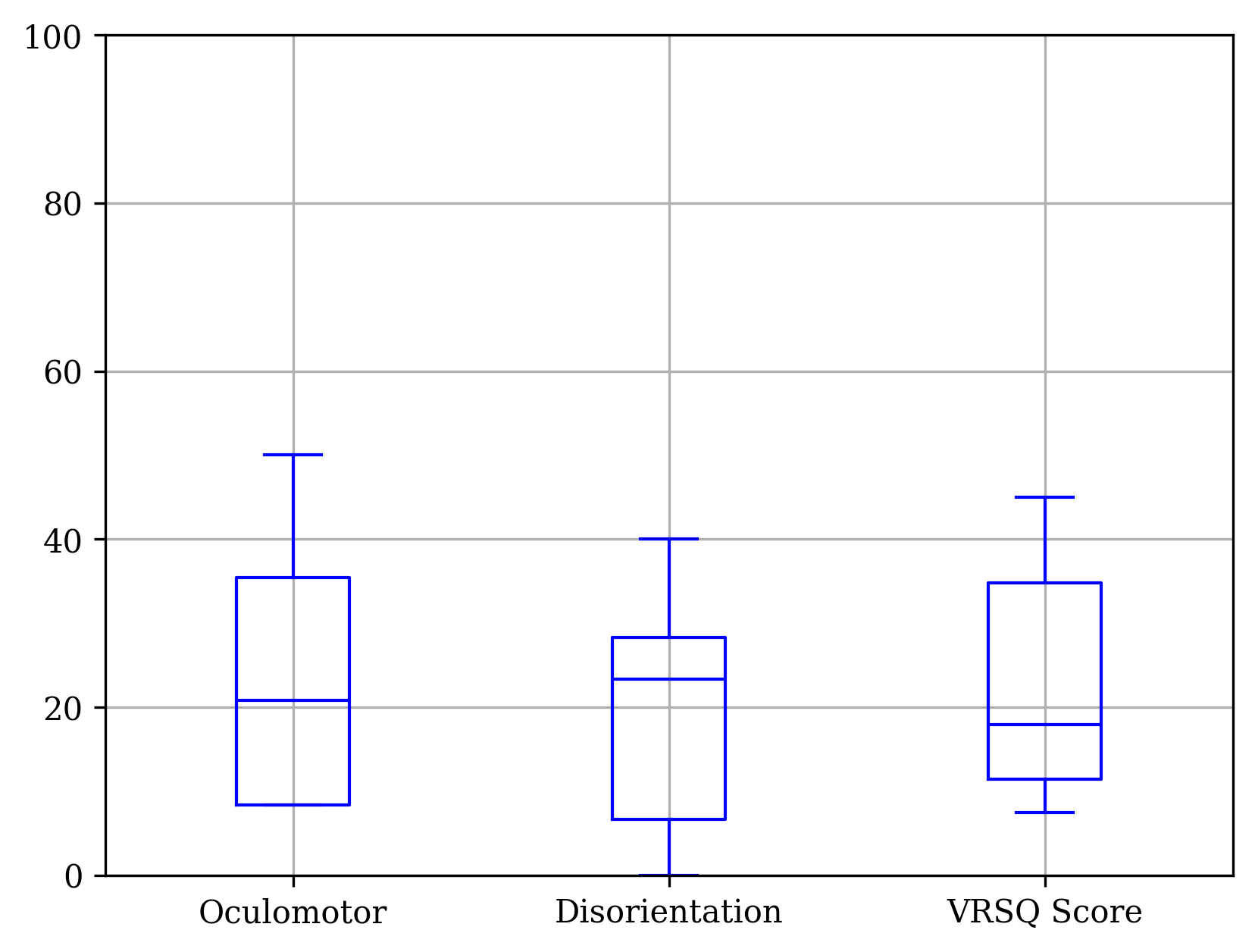}
  \caption{VRSQ scores of all VR participants}
  \Description{VRSQ scores of all VR participants incl. subscales as boxplots}
  \label{fig:result_vrsq_all_box}
\end{figure}

\subsubsection*{6) Motion Sickness}
To measure the motion sickness in \emph{SecCityVR} participants filled out the VRSQ. The rounded median scores of the dimensions are 20.83 for \emph{Oculomotor}, 23.33 for \emph{Disorientation}, and 17.92 for the \emph{VRSQ Score} with maximum scores of 50, 40, and 45 out of 100. As Figure~\ref{fig:result_vrsq_all_box} sets the score in relation to the full scale it shows that the overall score is still at the lower end of experienced motion sickness. The VRSQ results are of significant result with the p-values: $p_{ocu}=0.046^*$, $p_{dis}=0.012^*$ and $p_{vrsq}=0.033^*$. An additional sickness feature that four participants mentioned when discussing additional feedback was their experience of some kind of fear of heights, either at the beginning of the study or when looking down at the city.

\subsubsection*{7) Discussion}
\label{sec:user_study:discussion}
In the first study, dashboard users completed tasks in a significantly shorter median time, consistent with our expectations. This may be attributed to the dashboard's ability to always display package and class names, reducing the manual effort required to solve tasks T2 and T3, and to the novelty of VR-based reporting, which may have influenced user performance. On the other hand, the VR displays call graph information, which helps with T4, where dashboard users could have a disadvantage using the raw textual call graph. This fact could additionally explain why participants finished the fourth task faster in VR. However, the main impact on the time is the travel time in VR. This shows slightly in T2 where users mostly knew where they wanted to go to find the answer but they had to fly the full distance in VR and therefore were 26 seconds slower. The same is true for T3. Participants had limited the potential locations to the ends of connected call graph edges, but if it was not the graph node they were looking for, they had to backtrack or rise above the city to navigate further. On the dashboard, there is less travel time, the only limiting factor is the scroll and reading speed of users when searching a package by name or size. Despite differences in task completion times, all participants finished within the set time limits, so it can be concluded that the tools were effective in assisting the participants when solving the tasks but dashboard users were faster than the VR group. In the collaboration study, using the full time wasn’t due to tool limitations but rather the task design. Notably, one VR group and one dashboard group finished early, indicating that completion time was influenced more by the task and provided information than by the tool itself. An explanation for the early finish of half the study groups could be the experienced frustration measured by the TLX questionnaire and their additional feedback, more on this later in this section. As well as the task design itself: participants decided only to fix a vulnerability in the owner's application code but not in the dependencies.

Results show that a great majority of tasks were solved correctly by the VR users, except for some errors in the first task. They understood the assignments and were able to find a solution with an overall low error rate. Most VR participants struggled with the first task and instinctively tried to find the biggest class, not the package, which suggests that the code city, designed with classes as buildings, may be more suitable for an exploration task with a focus on classes and less on packages. The overall task correctness allows the conclusion that the hypothesis $H_1$ can be accepted and the $RQ_1$ can be answered with \emph{'software vulnerabilities can be effectively visualized by a VR code city including classes as buildings with methods as color-coded building floors containing the information about vulnerabilities and connections through directed arcs representing edges in the call graph'}. 

Participants provided a lot of constructive feedback which could help to improve the initial version of \emph{SecCityVR} and none of them really disliked the VR experience. As shown in the SUS questionnaire score, the VR environment achieved a median score of 70 across all users (single and collaboration). The dashboard was only rated with a median score of 58.75 across users. According to the analysis, this rates the VR environment as above average usability while the dashboard is rated as an 'Okay' score~\cite{SUS_Benchmarks}.

An unexpected result emerged from the TLX scores. While overall medians were similar between the dashboard and VR groups, two questions stood out: \emph{'How physically demanding was the task?'} and \emph{'How insecure, discouraged, irritated, stressed, and annoyed were you?'}. The median scores for these show a difference of 32.5 and 35 both indicating that the same tasks felt more rushed and stressed when done using a dashboard instead of VR. These results in the dimensions of temporal demand and frustration are mirrored in the collaboration study. This time the results for the temporal demand question differed even more. With a difference of 52.5, the dashboard group experienced that the task was rushed. The frustration kept a similar difference with 32.5 more points in the dashboard group. The reported similar effect in both studies of experiencing the pace of the tasks as rushed and being frustrated in the tasks is surprising. Especially because it was attempted to present all tasks the exact same for every participant, the tasks were read off a prepared experiment task list with the same wording every time. Therefore, the results could indicate an advantage of a VR environment over traditional tools with regard to the experienced motivation of users and being more relaxed when solving tasks. Considering that the dashboard users solved the tasks in a shorter time, an assumption can be made that when using a dashboard users feel stressed about the performance in a task because the displayed information is less engaging while in a VR environment users are more curious about the visualization and less focused on the tasks itself. Combining these findings with the higher calculated SUS for the VR environment the hypothesis $H_2$ could be accepted and interpreted as the \emph{SecCityVR} seems to have an overall better usability than the comparison dashboard. However, the reported result of the SUS questionnaire is not statistically significant according to the performed test, which leads to not accepting the hypothesis $H_2$. Therefore, $RQ_2$ is answered with: \emph{'participants felt a measurable improvement of lower temporal demand and lower frustration when using VR over a tabular dashboard, although they spent significantly more time overall to solve the same tasks in VR'}.

A factor that could highly influence the usability of the VR application is experiencing motion sickness when using it. The VRSQ scores result in a median reported motion sickness of lower than experienced \emph{slight} symptoms. In the collected feedback on motion sickness through the VRSQ no participant reported a symptom as \emph{severe}. However, when reviewing and comparing the given answers of the main study and the collaboration study, a trend is noticeable. In the collaboration setting, participants used the environment for a longer time and generally moved more through the city. It shows that the longer the usage of the VR environment, the higher the motion sickness symptoms. The median of all calculated VRSQ scores from the collaboration study corresponds to a rating in the middle of \emph{slight} to \emph{moderate} on the original questionnaire scale. Although there are some symptoms reported, these seem not to influence usability, most participants focused their additional feedback on controls and additional features rather than on motion sickness.

In addition to the motion sickness as an opposing factor of usability, we measured the experienced feeling of 'being there', using the IPQ. Except for realism, all subscale medians exceed the average rating of 3, indicating participants felt a sense of presence in the VR environment, including spatial presence and involvement. Results show that presence achieved favourable results overall and therefore did not have a critical negative impact on usability. The main factor reducing presence is the realism subscale, as the VR setting is an abstract visualization of a source code repository rather than a hyper-realistic world. We argue, and participants' feedback agreed, that adding realistic elements like skyscraper textures or moving cars would distract from the core visualization. 

A conclusion that can be drawn from the participants' additional feedback is that a set of features and changes could improve the efficiency or the usability of the VR environment. To provide a more efficient work environment features for reducing the travel time or task time like highlighting vulnerabilities (as 'review later' or 'already reviewed'), embedding the real source code in an IDE-like way, and teleportation to far points could be valid improvements. Revisiting the button and controller mapping was advised by some participants, too. The multi-user study setting has brought up suggested improvements specific to the collaboration features. First to better see which vulnerability the other user is talking about it was suggested to add a pointer to the synchronized controllers of a user. Additionally, seeing the current field of view, or alternatively, a realistic head instead of a cube, of the other user was mentioned as helpful. This feedback can be extended with findings that occurred during the testing of this setting against a tabular dashboard using the same collaborative task. The positive effects were noticeable in significantly lower perceived temporal demand. Although participants preferred \emph{SecCityVR}, given the lower task load, no significant statement about the usability can be made. However, during the collaboration study, it was observed that both teams used \emph{SecCityVR} differently, relying on different collaboration strategies and features. One team first split up and explored the code city on their own before regrouping and taking turns using the guided review feature to show their individual findings. The other team split up in a similar fashion, but every time they found something interesting, they asked their partner to use the teleport button to talk about their finding immediately. It can be concluded that the implemented collaboration features supported the participants well.

\subsubsection*{8) Threats to validity}
\label{sec:user_study:threats}
To ensure the evaluation process is thoroughly examined, the user study is judged according to the factors of \emph{construct validity, internal validity, external validity, and reliability} taken from the framework of \citeauthor{yin2009case}~\cite{yin2009case}.

To achieve \emph{construct validity}, the effects of motion sickness were measured with its own questionnaire and tried to be removed from the possible influences on usability. With the measurement of the experienced task load in addition to the usability, it was tried to mitigate the risk of restricting the measurements to narrowly on only one dimension of user satisfaction during the study and opening up the evaluation to possible influences by the task design.

Regarding the \emph{internal validity} possible alternative explanations were collected. Firstly, the motivation of participants could hinder the results, as they were volunteering for the study, maybe they were more enthusiastic about the topic than the majority of developers or students. Additionally, volunteering with the expectation to use the VR environment because it was in the title of the study, only to realize they are assigned to the control group using the baseline dashboard, could lower the motivation to perform well in the study tasks. In addition to that, the kind of information displayed on the baseline dashboard could be criticized. While in VR, the analysis results are displayed as an aggregated visualization, in the baseline, they are displayed as the raw results in a tabular form. Which could be argued to not be a suitable comparison when only the visualization properties are compared. This comparison was selected to keep the provided analysis data (issues, colors, information) identical when doing the tasks. However, another comparison of VR versus 2D graphical aggregation should be done in the future to rule out this threat. 

To address the \emph{external validity}, diverse groups were invited to mitigate the risk of excluding a potential user group and risking a threat to the generalization. Every group of participants had a different special interest, background, and set of skills. However, if a possible future user group does not fit these criteria this study may not apply to them. Statistical tests were done to ensure that the assumptions taken out of the results were significant or to mark them as not trustworthy. However, the amount of participants in the study is low, and statistical significance could be hard to reach or be error-prone. Therefore the results should be taken carefully and not be generalized without further validation. Due to organizational reasons, only industry participants were paired with industry participants while participants from the university and research background were paired with each other. No conclusions about mixed teams should be taken from this study.

In order to ensure the \emph{reliability} during the user study, a fixed order of execution was used every time. However, the different experiences with VR technologies could limit this factor. An initial tutorial in \emph{SecCityVR} was provided to minimize these risks.

\section{Conclusion and Future Work}
\label{ch:conclusion}
In conclusion, this paper demonstrates the potential of VR as a medium for visualizing and exploring software security vulnerabilities. The created proof-of-concept application, \emph{SecCityVR}, extends the code city metaphor into an immersive, collaborative environment that not only enhances spatial awareness of vulnerabilities but also facilitates real-time interaction among developers. The user study findings suggest that while task completion times were longer, the improved usability, reduced cognitive load, and lower frustration levels highlight the benefits of a VR approach over traditional dashboards. The results contribute to the fields of collaborative and secure software engineering, as well as software visualization. It provides a new application of VR code cities to visualize security vulnerabilities, as well as a novel environment for security audits using collaborative and immersive technologies.

Future improvements to the \emph{SecCityVR} environment include enhancing usability through features such as teleportation and source code integration. Collaboration could benefit from more realistic avatars and attention markers to improve communication. There are also plans to extend the proof-of-concept to support languages beyond Java. A promising direction for future research is exploring new perspectives of vulnerabilities with the code city visualization in VR. Shifting the focus from all vulnerabilities to a single vulnerability, a specific analysis technique, or concurrent vulnerabilities could provide valuable insights. Integrating existing ideas from static code analysis has several applications in VR and is encouraged to be tried out. Furthermore, researchers will be enabled to communicate their findings, algorithms, or techniques with the additional medium of VR.
\vspace{2cm}
\bibliographystyle{ACM-Reference-Format}
\bibliography{sample-manuscript}

\appendix

\end{document}